\documentclass[conference]{IEEEtran}
\IEEEoverridecommandlockouts
\usepackage{cite}
\usepackage{float}
\usepackage{amsmath,amssymb,amsfonts}
\usepackage{algorithmic}
\usepackage{graphicx}
\usepackage{textcomp}
\usepackage{xcolor}
\def\BibTeX{{\rm B\kern-.05em{\sc i\kern-.025em b}\kern-.08em
    T\kern-.1667em\lower.7ex\hbox{E}\kern-.125emX}}

\usepackage[colorlinks=true, allcolors=blue]{hyperref}
\usepackage{multirow}
\usepackage{array}
\usepackage{pifont}
\usepackage{listings}
\usepackage[hyphenbreaks]{breakurl}

\begin{document}

\title{ESDD2: Environment-Aware Speech and Sound Deepfake Detection Challenge Evaluation Plan}

\author{
Xueping Zhang$^{1}$, Han Yin$^{2}$, Yang Xiao$^{3}$, Lin Zhang$^{4}$, Ting Dang$^{3}$, \\ Rohan Kumar Das$^{5}$, Ming Li$^{1}$ \\
$^{1}$Duke Kunshan University 
$^{2}$Korea Advanced Institute of Science and Technology, \\
$^{3}$The University of Melbourne, 
$^{4}$Johns Hopkins University,
$^{5}$Fortemedia Singapore\\
esdd2-challenge@googlegroups.com
}

\maketitle

\begin{abstract}
Audio recorded in real‑world environments often contains a mixture of components:  foreground speech and background environmental sounds. With rapid advances in text‑to‑speech, voice conversion, and other generation models, either component can now be modified independently. 
Such component-level manipulations are harder to detect, as the remaining unaltered component can mislead the systems designed for whole deepfake audio, and they often sound more natural to human listeners.
To address this gap, we have proposed CompSpoofV2 dataset\footnote{CompSpoofV2: \url{https://xuepingzhang.github.io/CompSpoof-V2-Dataset/}} and a separation-enhanced joint learning framework\footnote{Baseline: \url{https://github.com/XuepingZhang/ESDD2-Baseline}}. CompSpoofV2 is a large-scale curated dataset designed for component-level audio anti-spoofing, which contains over 250k audio samples, with a total duration of approximately 283 hours. Based on the CompSpoofV2 and the separation-enhanced joint learning framework, we launch the Environment-Aware Speech and Sound Deepfake Detection Challenge (ESDD2)\footnote{ESDD2: \url{https://sites.google.com/view/esdd-challenge/esdd-challenges/esdd-2/description}}, focusing on component-level spoofing, where both speech and environmental sounds may be manipulated or synthesized, creating a more challenging and realistic detection scenario. The challenge will be held in conjunction with the IEEE International Conference on Multimedia and Expo 2026 (ICME 2026)\footnote{ICME 2026: \url{https://2026.ieeeicme.org/}}.

\end{abstract}


\section{Introduction}
In real-world recording scenarios, audio often contains a mixture of foreground speech and background sounds. We define such audio as comprising two components: (i) speech, i.e., linguistically meaningful speech produced by the primary foreground speaker, and (ii) environmental sound, i.e., any non‑speech background or non‑target speech. With rapid advances in text‑to‑speech, voice conversion, and other generation models, either component can now be modified independently. For example, replacing the background while leaving the foreground human speech unchanged, or modifying the speech content while preserving the background.
Such component-level manipulations are harder to detect, as the remaining unaltered component can mislead the systems designed for whole deepfake audio, and they often sound more natural to human listeners.

To address this gap, we introduced a component-level spoofing audio dataset CompSpoof~\cite{zhang2025compspoof} in 2025. Then, we extend the dataset to \textbf{CompSpoofV2}, which contains more than 250,000 audio clips (around 283 hours) formed by mixing bona fide (genuine or real) and spoofed (synthetic or manipulated) audio, including speech and environmental sounds, from multiple sources. We also proposed a \textbf{separation-enhanced joint learning framework} to explore to address this gap.

To further promote the development of environment-aware speech and
sound deepfake detection, we extend our first ESDD challenge\cite{ESDD1} and launch the ICME 2026 Environment-Aware Speech and Sound Deepfake Detection Challenge (ESDD2). While our first ESDD challenge focuses on the detection of forged environmental sounds, ESDD2 advances this direction by addressing component-level spoofing, where both speech and environmental sounds may be manipulated or synthesized, posing a more challenging and realistic detection scenario.

\section{Dataset}

\begin{figure*}[!h]
\centerline{\includegraphics[width=1\textwidth]{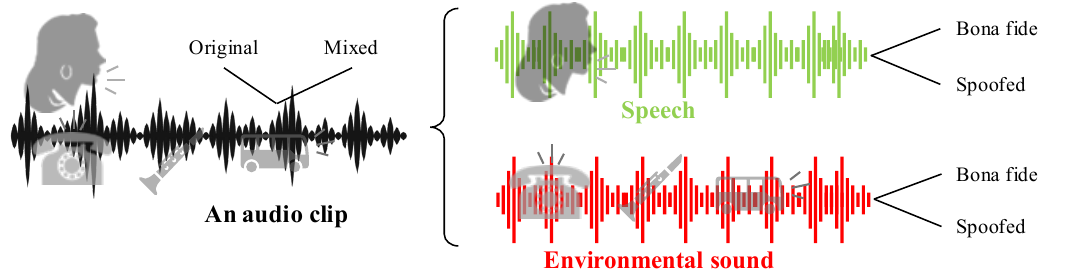}}
\caption{ESDD2 Task illustration. An audio clip is first classified as mixed or original; for mixed audio, speech and environmental sound components are separately evaluated for genuineness, resulting in five target classes.}
\label{main}
\end{figure*}

The \textbf{CompSpoofV2} is a dataset designed for component-level anti-spoofing detection research, where either the speech or the environmental sound component (or both) may be spoofed.
CompSpoofV2 contains over 250,000 audio samples, with a total duration of approximately 283 hours. Each audio sample has a fixed length of 4 seconds and is provided at multiple sampling rates, enabling a more faithful simulation of real-world acoustic and system-level variations.
Building upon the CompSpoof dataset\cite{zhang2025compspoof}, CompSpoofV2 significantly expands the diversity of attack sources, environmental sounds, and mixing strategies. In addition, newly generated audio samples are distributed across the test set and are specifically designed to serve as detection data under unseen conditions.

In CompSpoofV2, each audio belongs to one of five categories, covering all combinations of bona fide and spoofed speech and environmental sound as shown in Table \ref{tab:CompSpoof_classes}. \textbf{The task of this challenge is to classify a given audio clip into one of the five defined classes}, as shown in Figure \ref{main}. The audio sources for each category are summarized in Table~\ref{tab:audio_sources_train} and Table~\ref{tab:audio_sources_eval}.

\begin{table*}[h]
\caption{Five Categories Considered in CompSpoofV2 dataset.}
{
\centering{
\begin{tabular}{ c c c c c l}
\hline
\textbf{ID} & \textbf{Mixed} & \textbf{Speech} & \textbf{Environment} & \textbf{Class Label} & \textbf{Description} \\
\hline
0 & \ding{55} & Bona fide & Bona fide &original & Original audio without any manipulation or mixing\\
1 & \ding{51} & Bona fide & Bona fide &bonafide\text{\_}bonafide & Bona fide speech mixed with bona fide environmental sound from other audio\\
2 & \ding{51} & Spoofed   & Bona fide &spoof\text{\_}bonafide & Spoofed speech mixed with bona fide environmental sound \\
3 & \ding{51} & Bona fide & Spoofed   &bonafide\text{\_}spoof & Bona fide speech mixed with spoofed environmental sound \\
4 & \ding{51} & Spoofed   & Spoofed   &spoof\text{\_}spoof & Spoofed speech mixed with spoofed environmental sound \\
\hline
\end{tabular}
}
}
\label{tab:CompSpoof_classes}
\end{table*}

\begin{table*}[t]
\centering
\caption{Audio sources for the training and validation sets.}
\label{tab:audio_sources_train}
\setlength{\tabcolsep}{2pt} 
{\fontsize{7.2pt}{10pt}\selectfont
\begin{tabular}{l|@{}ccc@{}}
\hline
\textbf{ID} & \textbf{Original Source} & \textbf{Speech Source} & \textbf{Environmental Sound Source} \\
\hline
0 & AudioCaps\cite{AudioCaps}, VGGSound\cite{VGGSound} & -- & -- \\
1 & -- & CommonVoice\cite{CommonVoice}, LibriTTS\cite{LibriTTS}, english-conversation-corpus\cite{ecc} & AudioCaps, TAUUAS\cite{TAUUAS}, TUTSED\cite{TUTSED2016, TUTSEDdev2017, TUTSEDeval2017}, UrbanSound\cite{UrbanSound}, VGGSound \\
2 & -- & CommonVoice, LibriTTS & EnvSDD\cite{EnvSDD}, VcapAV\cite{VcapAV} \\
3 & -- & ASV5\cite{ASV5}, MLAAD\cite{MLAAD} & AudioCaps, TAUUAS, TUTSED, UrbanSound, VGGSound \\
4 & -- & ASV5, MLAAD & EnvSDD, VcapAV \\
\hline
\end{tabular}
}
\end{table*}

\begin{table*}[t]
\centering
\caption{Audio sources for the evaluation and test sets.}
\label{tab:audio_sources_eval}
\begin{tabular}{l|ccc}
\hline
\textbf{ID} & \textbf{Original Source} & \textbf{Speech Source} & \textbf{Environmental Sound Source} \\
\hline
0 & AudioCaps, VGGSound & -- & -- \\
1 & -- & CommonVoice, LibriTTS, english-conversation-corpus & AudioCaps, TAUUAS, TUTSED, UrbanSound, VGGSound \\
2 & -- & CommonVoice, LibriTTS & EnvSDD, VcapAV, \textbf{New Generated} \\
3 & -- & ASV5, MLAAD, \textbf{New Generated} & AudioCaps, TAUUAS, TUTSED, UrbanSound, VGGSound \\
4 & -- & ASV5, MLAAD, \textbf{New Generated} & EnvSDD, VcapAV, \textbf{New Generated} \\
\hline
\end{tabular}
\end{table*}

The training and validation sets share the same data sources and class distribution.
Similarly, the evaluation and test sets share the same data sources and class distribution.
Notably, the evaluation and test sets include newly generated audio samples that are unseen in the training and validation sets.

\section{Evaluation Criteria}

System performance will be evaluated using the \textbf{Overall Macro-F1 score} across all five target classes, as shown in ~\eqref{eq1}. The \textbf{Macro-F1 score} is computed as the arithmetic mean of the per-class F1 scores. This ensures all classes contribute equally to the final evaluation, regardless of sample imbalance.
\begin{equation}
\label{eq1}
\text{Macro-F1} = \frac{1}{5} \sum_{i=1}^{5} \text{F1}_i, 
\end{equation}
where, the $\text{F1}_i$ score is calculated as the harmonic mean of precision $P_i$ and recall $R_i$ for class $i$:
\begin{equation}
\label{eq2}
\text{F1}_i = \frac{2 \cdot P_i \cdot R_i}{P_i + R_i}, \quad
P_i = \frac{\text{TP}_i}{\text{TP}_i + \text{FP}_i}, \quad
R_i = \frac{\text{TP}_i}{\text{TP}_i + \text{FN}_i}.
\end{equation}
$\text{TP}_i$, $\text{FP}_i$, and $\text{FN}_i$ denote the number of true positives, false positives, and false negatives for class $i$, respectively.
A higher Macro-F1 score indicates better overall system performance.

In addition, we introduce three auxiliary metrics based on equal error rate (EER): $\mathrm{EER}_{\text{original}}$, $\mathrm{EER}_{\text{speech}}$ and $\mathrm{EER}_{\text{env}}$.
$\mathrm{EER}_{\text{original}}$ is the EER for distinguishes the \texttt{Original} ($i=0$) class from all other four classes $i=\{1,2,3,4\}$. Similarly, $\mathrm{EER}_{\text{speech}}$ is the EER for detecting spoofed speech component, while $\mathrm{EER}_{\text{env}}$ is the EER for detecting spoofed environmental component. The three EER metrics are included for diagnostic and analysis purposes only and will \textbf{not} be considered for the leaderboard ranking.


\section{Baseline}
We propose a \textbf{separation-enhanced joint learning framework}~\cite{zhang2025compspoof} as the baseline for this task. The framework first detects potentially spoofed mixtures, then separates the audio into speech and environmental components, which are then processed by speech- and environment-specific anti-spoofing models. Their outputs are fused and mapped to five-class predictions. The separation and anti-spoofing models are jointly trained to preserve spoofing-relevant cues. The baseline results are as \ref{tab:performance}.
\begin{table}[H]
\centering
\caption{Performance (EER \% and F1-score) of baseline model on the validation, evaluation, and test sets of CompSpoofV2.}
\begin{tabular}{l|cccccc}
\hline
          & F1-score & Original EER & Speech EER & Env. EER   \\ \hline

Val. set & 0.9462  & 0.0031 & 0.0172 & 0.3766  \\
Eval set & 0.6224  & 0.0174 & 0.1993 & 0.4336  \\
Test set & 0.6327  & 0.0173 & 0.1978 & 0.4279   \\ \hline
\end{tabular}
\label{tab:performance}
\end{table}

\section{Submission Guideline}
We use CodaBench platform for results submission\footnote{Results Submission: \url{https://www.codabench.org/competitions/12365/}}.
Participants are expected to compute and submit a text file as follows, in which each line contains five columns.
From left to right, the columns include:
\begin{itemize}
 \item The audio ID;
 \item The predicted class ID, which is an integer label from 0 to 4 corresponding to the class definitions in Table~\ref{tab:CompSpoof_classes};
 \item The confidence score for predicting the original class;
 \item The confidence score for the speech component;
 \item The confidence score for the environmental sound component.
\end{itemize}
\begin{lstlisting}
  UGBFJ3.wav |  0  |  1.22  |  0.91  |  0.99 
  ABCD12.mp3 |  3  | -1.54  |  0.89  | -1.89 
\end{lstlisting}

Submitting the scores is not mandatory. If the participant's method does not generate the scores, they \textbf{MUST} use a hyphen (-) as a placeholder in the last three columns of the file as example shown in below:
\begin{lstlisting}
  UGBFJ3.wav |  0  |  -  |  -  |  - 
  ABCD12.mp3 |  3  |  -  |  -  |  - 
\end{lstlisting}

The challenge has two phases: the preparation phase and the final ranking phase. During the preparation phase, participants submit evaluation set predictions and the leaderboard reflects evaluation set rankings. During the final ranking phase, participants submit test set predictions and the leaderboard determines the final ranking. Each team may submit up to ten results in the final ranking phase, with the best three used for final ranking to ensure fairness. After the competition, we will release the metadata of the evaluation and test sets to support reproducibility and future research.

Participants may use additional public datasets only with the organizers’ prior approval. Requests must be submitted by email to the organizers (xueping.zhang@dukekunshan.edu.cn) before February 20th, 2026. The organizers will publish the list of approved datasets; any dataset not on this list is not permitted for training.

\section{Rules}

To ensure fairness, the following rules are established:
\begin{itemize}
    \item The evaluation and test sets must \textbf{NOT} be used for training.
    \item Publicly available pre-trained models released prior to January 1, 2026 are allowed but must be disclosed.
    \item Participants are encouraged to report model parameters and computing resources.
    \item Data augmentation is permitted. But the use of any additional synthetic audio--including audio generated by Text-to-Speech models, voice conversion models, or any other audio generation models--is strictly prohibited.
    \item Participants may use additional public datasets only with the organizers' prior approval. Requests must be submitted to the organizers (xueping.zhang@dukekunshan.edu.cn) before February 20, 2026, and the organizers will publish the list of approved datasets; any datasets not on this list are not permitted for training.
\end{itemize}

\section{Registration Process}
The following Google Form will be used by participating teams for registration of their respective teams in the challenge\footnote{Registration: \url{https://forms.gle/WnkDkyiXtoVNMZRk7}}.
\section{Important Dates}
The detailed timeline is shown in Table \ref{timeline}.

\begin{table}[!h]
\centering
\caption{ESDD2 Challenge Timeline.}
\begin{tabular}{p{5.5cm}|p{2.3cm}}
\hline 
Registration Opens & January 10th, 2026 \\ \hline
Training, validation and evaluation sets release; Baseline code release; Preparation-phase leaderboard update & January 30th, 2026 \\ \hline
Additional dataset approval deadline &  February 20th, 2026 \\ \hline
Test set release; Final ranking phase leaderboard update & March 20th, 2026 \\ \hline
Registration closes, leaderboard freeze and Grand Challenge Results Notification& April 25th, 2026 \\ \hline
Challenge Paper Submission Deadline & April 30th, 2026 \\ \hline
Challenge Paper Acceptance Notification & May 7th, 2026 \\ \hline
Challenge Paper Camera-Ready Deadline & May 15th, 2026 \\ \hline
\end{tabular}
\label{timeline}
\end{table}
More information can be found in the website (\href{https://sites.google.com/view/esdd-challenge/esdd-challenges/esdd-2/description}{https://sites.google.com/view/esdd-challenge/esdd-challenges/esdd-2/description})

\section{Challenge Papers Submission}
The ESDD2 challenge is one of the Grand Challenges in ICME 2026. Challenge papers for Grand Challenges must follow the same formatting standards as ICME 2026 regular conference papers. Submissions are limited to 6 pages, including all text, figures, and references. Detailed author instructions can be found here: ICME 2026 Author Information (\url{https://2026.ieeeicme.org/author-information-and-submission-instructions/}). Upon acceptance, the challenge papers will be published in the conference workshop proceedings.

\section{Sponsors and Awards}
OfSpectrum, Inc. (\url{https://ofspectrum.com/}), an AI company based in the U.S., is sponsoring this challenge and will provide 1,000 U.S. dollars prize to the first-place winner. OfSpectrum develops imperceptible watermarking technology, with applications in content provenance, copyright protection and facilitating the safe and legitimate use of AI in voice applications.

\bibliographystyle{IEEEbib}
\bibliography{icme2025references}


\end{document}